\ifx\mnmacrosloaded\undefined \input mn\fi
\input epsf
\pageoffset{-2.0pc}{0pc}

\def\n{$n$ }
\def\eps{$\epsilon$}
\def\eps1{$\epsilon$ }
\def\rq{$R^{1/4}\ $}
\def\rqe{$R^{1/4}$}
\def\rh{$R^{1/2}\ $}
\def\rhe{$R^{1/2}$}

\def\en{{\cal E }}
\def\Reff{R$_{\rm eff}$}
\def\Reffs{R$_{\rm eff}$ }
\def\efen{{\it f(\cal E) }}
\def\solarm{M$_\odot$}
\def\tentensol{$10^{10}$M$_\odot$ }
\def\hd{h$_{\rm d}$}
\def\Md{M$_{\rm d}$}
\def\hds{h$_{\rm d}$ }


\begintopmatter

\title {The effects of a disc field on bulge surface brightness}

\author{Y. C. Andredakis}

\affiliation{Kapteyn Astronomical Institute, P.O. Box 800, 9700 AV Groningen,
             The Netherlands (jandr@astro.rug.nl) } 

\shortauthor{Y. C. Andredakis}
\shorttitle{Effects of disc formation on the bulge}

\abstract {Collisionless N-body simulations are used in an effort to
reproduce the observed tendency of the surface brightness profile of bulges
to change progressively from an \rq law to an exponential, going from early to
late type spirals. A possible cause for this is the formation of the disc,
later in the history of the galaxy, and this is simulated by applying on the
N-body bulge the force field of a exponential disc whose surface density
increases with time. It is shown that $n$, the index of the 
Sersic law $\Sigma_n(r) \propto \exp[-(r/r_0)^{1/n}] $
that best describes the surface brightness profile, does indeed decrease from 4
(de Vaucouleurs law) to smaller values; this decrease is larger for more
massive and more compact discs. A large part of the observed trend of $n$
with B/D ratio is explained, and many of the actual profiles can be matched
exactly by the simulations. The correlation between the disc scalelength and
bulge effective radius, used recently to support the ``secular evolution"
origin for bulges, is also shown to arise naturally in a scenario like this.
This mechanism, however, saturates at around $n=2$ and exponential
bulges cannot be produced; as $n$ gets closer
to 1, the profile becomes increasingly robust against a disc field. These
results provide strong support to the old-bulge hypothesis for the early-type
bulges. The exponential bulges, however, remain essentially unexplained; the
results here suggest that they did not begin their lives as \rq spheroids,
and hence were probably formed, at least in part, by different processes than 
those of early type spirals.}


\maketitle
		
\section{Introduction}

The question of when the bulges of spiral galaxies formed and by which
mechanism, remains open. We can distinguish in general two opposing ideas on
the subject. According to the first one, which might be called the
traditional picture, the bulge formed well before the disc of the galaxy, 
as the first stage of the collapse of a galaxy-sized density perturbation. 
This is advocated by the ELS scenario for galaxy formation (Eggen, Lynden Bell 
\& Sandage, 1962) and its modern revisions, such as bulge formation through 
infall of enriched gas from the star-forming halo (Carney et al. 1990). 
Other ``early bulge'' scenarios are the ones by
Kauffmann, White \& Guiderdoni (1993) and Baugh, Cole \& Frenk (1996), in which
bulges form by merging of stellar discs inside hierarchically merging dark 
halos. In this last scheme the disc that we see today is assembled later by 
gas accretion. 

These scenarios are supported by a body of evidence regarding the age of
bulges. The main tracers of the bulge population are as a rule old objects such
as RR Lyrae stars, planetary nebulae and globular clusters. Based on age
determinations using stellar populations, the bulge of the Milky Way galaxy
seems to be as old as the oldest globular clusters (Ortolani et al. 1995).
Metallicity studies by Jablonka et al. (1996) favour a rapid early formation
for a sample of 28 extragalactic bulges. Finally, bulge velocity fields
(Kormendy \& Illingworth 1982), and the high central surface brightness of
bulges also point toward a formation by dissipational collapse, before the
stellar disc or perhaps even the halo was formed.

On the other hand, it has been shown that a bulge can be formed after the disc,
through secular evolution phenomena. Numerical experiments (Combes \& Sanders
1981, Combes et al. 1990, Norman, Sellwood \& Hasan 1996) have shown that
bulges can be created by dissipationless processes such as the thickening or
destruction of a bar, and many bulges display the characteristics expected in
such formation scenarios, such as triaxial or peanut shape, cylindrical
rotation etc (Shaw 1993). Furthermore, there does not appear to be a
significant difference between disc and bulge in terms of colours (Balcells \&
Peletier 1994) and many bulges display disc-like kinematics (Kormendy 1992).

If the disc is the one that formed later, it is quite probable that its
formation has left a signature on some of the properties of the pre-existing
bulge. An indication of such a signature may be the surface brightness profile
of the bulge, as considered by Andredakis, Peletier and Balcells (1995),
hereafter APB95.  In general bulges can be described by laws belonging to the
family $\Sigma_n(r) \propto \exp[-(r/r_0)^{1/n}] $ (Sersic 1968) where
$\Sigma(r)$ is the surface brightness as a function of radius, 
$r_0$ is a general scalelength and $n$ can take any
positive value. APB95 found that the surface brightness profile has a different
shape depending on the morphological type of the galaxy and---subsequently---on
the bulge-to-disc ratio; more specifically, that the index $n$ of Sersic's law
changes smoothly and systematically from values of 4 or bigger down to 1 as we
go from early to late-type spirals. This effect has since been confirmed by de
Jong (1995) and Courteau, de Jong and Broeils (1996) on top of the indications
that already existed (Frogel 1988, Kent, Dame \& Fazio 1991, Andredakis \&
Sanders 1994). Recently, indications of this behaviour were found also in the
kinematics of spirals (Heraudeau et al, 1996). The systematic change of \n was
interpreted by APB95 as an imprint of the disc formation and this implied that
the disc formed later than the bulge in all these different types of spirals.

In this paper, it is attempted to confirm by an N-body simulation of this
process whether this change in \n is indeed the signature of a disc forming
around the bulge. Assuming that the ``proto-bulge" is an equilibrium system
following the \rq law, a slowly strengthening disc field is introduced and the
effects (if any) that this has on the bulge surface brightness profile are studied.  The
paper is organized as follows: Section 2 describes the set-up of the N-body
models, the diagnostics used and the study of the inherent stability of the
models. The results of the evolution of a template model with a disc are presented
in detail in Section 3, and the effects of the presence of a dark halo are
examined. In Section 4 the results of the simulations are compared to the
observed trends, and the evolution of bulges with different initial density
laws is described. A short discussion of the bulge formation scenarios in 
light of the results in this paper is given in Section 5, and finally the
conclusions are given in Section 6.

It should be noted here that the response of an N-body bulge to the formation
of a disc has been studied before by Barnes \& White (1984), hereafter BW84,
but from a somewhat different point of view. They attempted to determine
whether bulges were more like ellipticals in the past, before the disc was
created; in this paper, the emphasis is put on trying to find traces of the
disc formation in present-day bulges.

\section{N-body realizations of bulges}

\subsection{Construction of the N-body models}

In the simulations described, any evidence of late formation of the late-type
bulges is for the moment ignored.  As mentioned in the previous Section, it
is assumed that the bulge is already in place when the disc begins to form,
and the results will test this hypothesis. It is further assumed that the
surface brightness profile of the ``proto-bulge'' follows an \rq law. It has
been shown (van Albada 1982, Theis \& Hensler 1993) that the collapse of
density clumps and the subsequent violent relaxation produces \rq law
spheroids, and the bulges of S0s, that are supposedly less affected by the
disc due to their greater mass, are very well represented by \rq laws.  The
conclusions do not depend heavily on this assumption, as other forms for the
initial shape of the profile will also be used.

The problem of constructing equilibrium models that follow a certain law
(\rqe, \rhe, exponential or anything else) is essentially one of calculating
the distribution function (DF) that self-consistently generates the required
surface brightness profiles. The calculations here are based on systems that
are initially isotropic and spherical, i.e. the DF is a function of energy
only, $f=\efen$. Rotating models are constructed by imposing a certain amount 
of rotation on the already existing isotropic models, following BW84.
Oblate models will not be considered in this study; constructing detailed, 
oblate rotating spheroids that follow a certain density law {\it ab initio}
is still a difficult computational problem and lies somewhat beyond the scope 
of this paper. 

For the particular purpose here, i.e. to examine changes in the surface brightness 
profile, the requirement is that the initial models should follow the \rq law 
exactly. There are some self-consistent models in the literature with
analytical expressions for the space density and the DF, which approximate
the \rq law in projection, such as the $\gamma_{3/2}$ model of Dehnen (1993)
(or $\eta_{3/2}$ in the notation of Tremaine et al., 1994). However, their
approximate nature introduces large uncertainties in the determination
of the best-fit $n$.
Therefore, the density of the models in this study is a deprojection of the 
\rq law. The density and the potential are calculated numerically and the DF
is subsequently calculated using the method of Binney
(1982). This allows the construction of spherical isotropic models with any
desired density law. Inverting the relation between the density $\rho(r)$ and
the DF $\efen$ (Eddington's formula), and then using Poisson's equation to 
substitute the derivatives of the density with respect to the potential, one 
finds that 
$$ \efen = {1\over \sqrt{8}\pi^2}\,\int_0^\en
           {d^2\rho\over d\Psi^2}
          {d\Psi\over \sqrt{\en - \Psi}} \eqno\stepeq $$
where
$$ {d^2\rho\over d\Psi^2} = 
          \Bigl({r^2\over GM}\Bigr)^2\,
          \Bigl[{d^2\rho\over dr^2} + {d\rho\over dr} \,
          \Bigl({2\over r} - {4\pi\rho\,r^2\over M}\Bigr)\Bigr]. $$
$\Psi$ and $M$ are the potential and the mass as a function of radius,
and $\en$ is the {\it relative} energy, defined as $\en = \Psi(r) - v^2/2$.
This expression is integrated numerically to obtain the distribution
function for the anticipated values of the relative energy $\en$. The construction
of the N-body model is then straightforward. Using the rejection method, a 
random radius is attributed to each particle such that the desired surface 
brightness is generated; then, a velocity is assigned to each particle 
according to the DF at that particular radius. 
In calculating the potential $\Psi$,
a softening parameter $\epsilon$ is introduced, whose value is equal to the 
softening length used 
in the N-body code for the calculation of the force (i.e. the potential
is no longer exactly Newtonian). The use of this potential softening
parameter in the building of the initial model helps to avoid effects like
the ``smearing" of central cusps in the density distribution that are common 
in N-body experiments (see for example the time evolution of isolated Hernquist
models in Navarro, Eke \& Frenk (1997) or the equivalent for Evans models in
Kuijken \& Dubinski (1994)).
Therefore, the expression for the potential that is used in equation (1) is
$$ \Psi(r) = {M(r)\over r+\epsilon} + 
             4\pi\int_r^\infty\!r\rho(r)dr \eqno\stepeq $$ 
where $M$ and $\rho$ are the mass and density of the bulge at radius $r$. 

Finally, rotating models are constructed by
introducing a term that involves the z-component of the angular momentum in
the isotropic DF. This rotating DF has the form
$$ f({\cal E},J_z) = f_{iso}~exp(\gamma J_z) \eqno\stepeq $$
where $f_{iso}$ is the DF of eq. (1).
This general form of DF was used fairly successfully by Jarvis and Freeman
(1985) to fit the bulge kinematics of edge-on spirals.  In effect, this
favours velocity distributions that are skewed toward high positive
$v_{\phi}$.  The N-body realization is constructed as before; the difference
is that since the DF is not odd in $J_z$, the integral of $f$ in velocity
space does not give back the initial density and the model is no longer
completely self-consistent. Therefore, the initial system is allowed to
relax for several dynamical times before applying the disc field. The result
for values of $\gamma \approx 1$ is still a spherical bulge that has
practically the same surface brightness and velocity dispersion profile but
with streaming velocities around the $z$-axis ($v_\phi$) that rise linearly 
with radius and reach a
constant value at around 1.5\Reffs (Here and in the rest of this paper, \Reffs
stands for the radius that contains half of the total number of particles,
in projection).

\subsection{Diagnostics and stability of the models}

The parameters of the bulge models of most interest
are the ellipticity, the effective radius and, of course, the index \n of 
the best-fitting power law of the surface brightness profile.
The ellipticity is determined as described in Dubinski and 
Carlberg (1991) and Katz (1991), by diagonalizing the inertia tensor
$$ M_{ij} = \sum {x_ix_j \over \alpha^2}, \hbox{\hskip 0.5truecm{\rm where}
            \hskip 0.5truecm} \alpha^2 = \bigl(x^2 + {y^2\over q^2} +
            {z^2\over s^2}\bigr). \eqno\stepeq $$ 
If we denote the length of the three principal axes of the bulge along the
x, y and z-directions with $a$, $b$ anc $c$ respectively, then $q=b/a$ and
$s=c/a$.
To compute the desired axis ratios $q$ and $s$ it is assumed that to a first
approximation they are both equal to 1. Then $M_{ij}$ is brought to diagonal 
form, and $q$ and $s$ are determined from
$$ q = \bigl({M_{yy}\over M_{xx}}\bigr)^{1/2} 
       \hbox{\hskip 0.5truecm{\rm and}\hskip 0.5truecm} 
   s = \bigl({M_{zz}\over M_{xx}}\bigr)^{1/2} \eqno\stepeq $$
These values are then substituted back into $M_{ij}$ and the procedure is
repeated until convergence to the desired tolerance is achieved. In this way
one has both the direction cosines of the principal axes (i.e. the position
angle) and the axis ratios with an accuracy of around 1--3\% for the 10000 to
30000 particles that comprise the models. From this point on, the 
axis ratio $c/a$ will be used in place of the ellipticity $\epsilon$
to show the changes in the shape of the bulge. The surface brightness profile 
is determined simply by looking at the model face-on (in case the flattening is
non-zero) and measuring the projected distribution of particles in circular
annuli. (For better statistics, a logarithmic spacing in radius is used.)
Knowing the total number of particles, the effective radius is then
immediately available. Finally, a function of the form $ \Sigma(r) =
\Sigma_0\exp[-(r/r_0)^{1/n}] $ is fitted to the profile, leaving all three
parameters ($\Sigma_0, r_0$ and $n$) free. In this way $n$ is determined, and we
have a consistency check between the (properly scaled) $r_0$ from the fit and
the directly determined half light radius. For the fit the points are
weighted by the Poisson error $1/\sqrt{N}$ where N is the number of particles
contained in each annulus. If an equal weighting is used instead, the
difference in the resulting $n$ is not more than 2\%.

\begintable{1}
\caption{{\bf Table 1.} Parameters used in the treecode}
{\halign{
\rm#\hfil&
\hskip 67pt\rm\hfil#\cr
\noalign{\vskip8pt}
Mass of the bulge       & 1. ($\times$\tentensol) \cr
Effective radius        & 1. (kpc) \cr
Gravity constant (G)    & 1 \cr
Particles               & 32768 \cr
Timestep                & 0.025 \cr
Total time              & 300 time units (1.5 Gyr) \cr
Opening angle $\theta$  & 1$^{\circ}$ \cr
Softening length        & 0.02 \cr
}}
\endtable
To evolve the models in time, an N-body treecode (Barnes \& Hut 1986) is used.
The parameters used in a typical run are given in Table 1.
The system of units used is M = \Reffs = 1. For the scale
of the systems studied here, it is convenient to attribute 1 kpc to the unit
of length and \tentensol to the unit of mass and they will be referred to as
such from now on.
The timestep (0.025) is chosen such that it is small compared to 
the bulge dynamical time at about $r=0.1$ kpc. The timestep is given by  
$dt \simeq \epsilon/\langle u^2 \rangle^{1/2}$, where $\langle u^2 \rangle$ 
is the velocity dispersion at $r=0.1$ kpc, and $\epsilon$ is the softening length. 
This is the minimum radius of interest in these systems at the moment; it would 
correspond to 0.5'' or, roughly, the size of one pixel for the distances of 
the galaxies studied in APB95. This radius (0.1 kpc) will be a lower 
limit in the subsequent calculations of the surface brightness profile 
parameters. Most of the calculations are done only to monopole accuracy in
the determination of the potential; a few test runs showed
that due to the high degree of symmetry of the system there is no perceptible
difference in the results if the (time-consuming) calculation of the
quadrupole terms is omitted.


\section{Evolution of models with a disc}

\subsection{Remarks on the evolution of a generic model}

Before applying any external field to the bulge models, it is 
necessary to ensure that the models are stable when left to evolve by
themselves. In general, one would expect these systems to be stable both
against radial and non-radial perturbations, since the distribution function
is everywhere positive and also $df/d\en > 0$ (Antonov 1962, Doremus et al.
1971). However, small spontaneous changes in the structure of the system, 
especially at
small radii, happen very often, mostly as a result of gridding or softening
effects. The initial \rq bulge is thus left to evolve in isolation for an 
amount 
of time equal to the one that used for the runs with the disc, and 
the evolution of the axis ratio, the effective radius and the index $n$ is
monitored. 
\beginfigure{1}
\epsfxsize=7.8cm \epsfbox{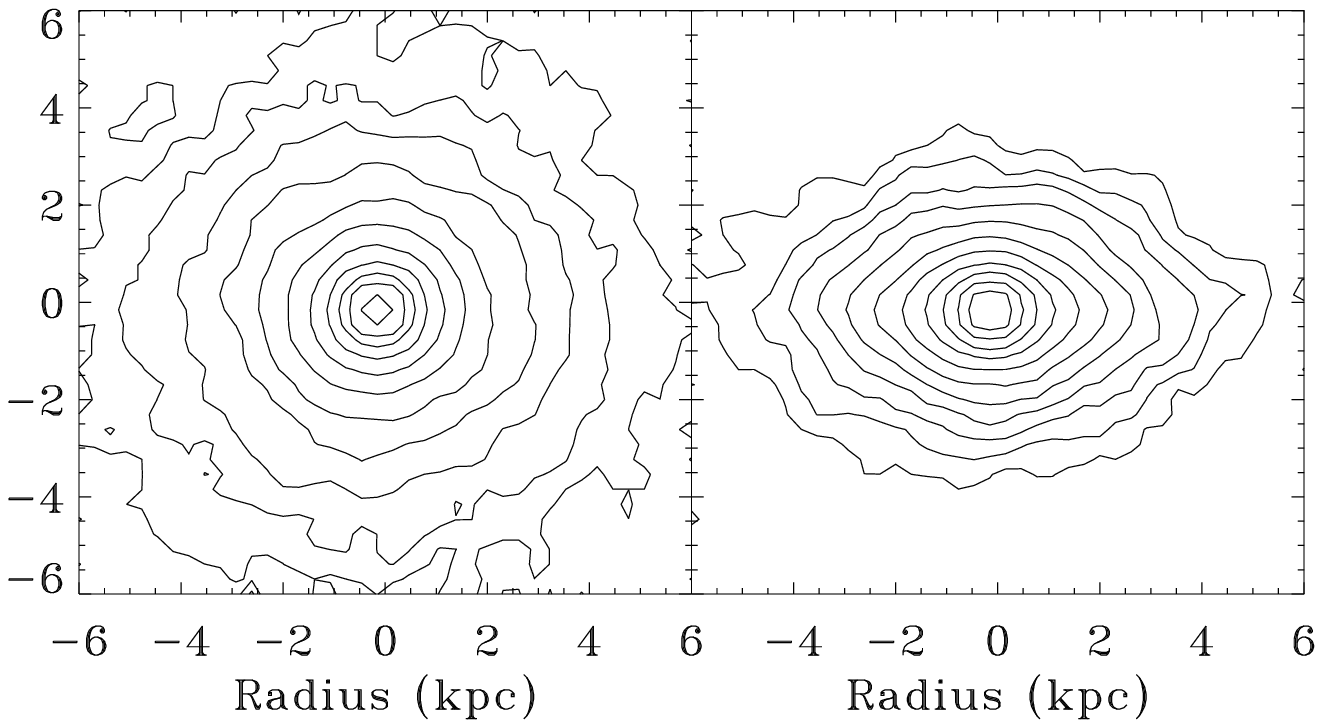}
\caption{{\bf Figure 1.} 
Contour plots of a typical bulge model seen edge-on, before and after the growth of the disc.
The contour levels are the same in both plots.}
\endfigure

\beginfigure{2}
\epsfxsize=7.8cm \epsfbox{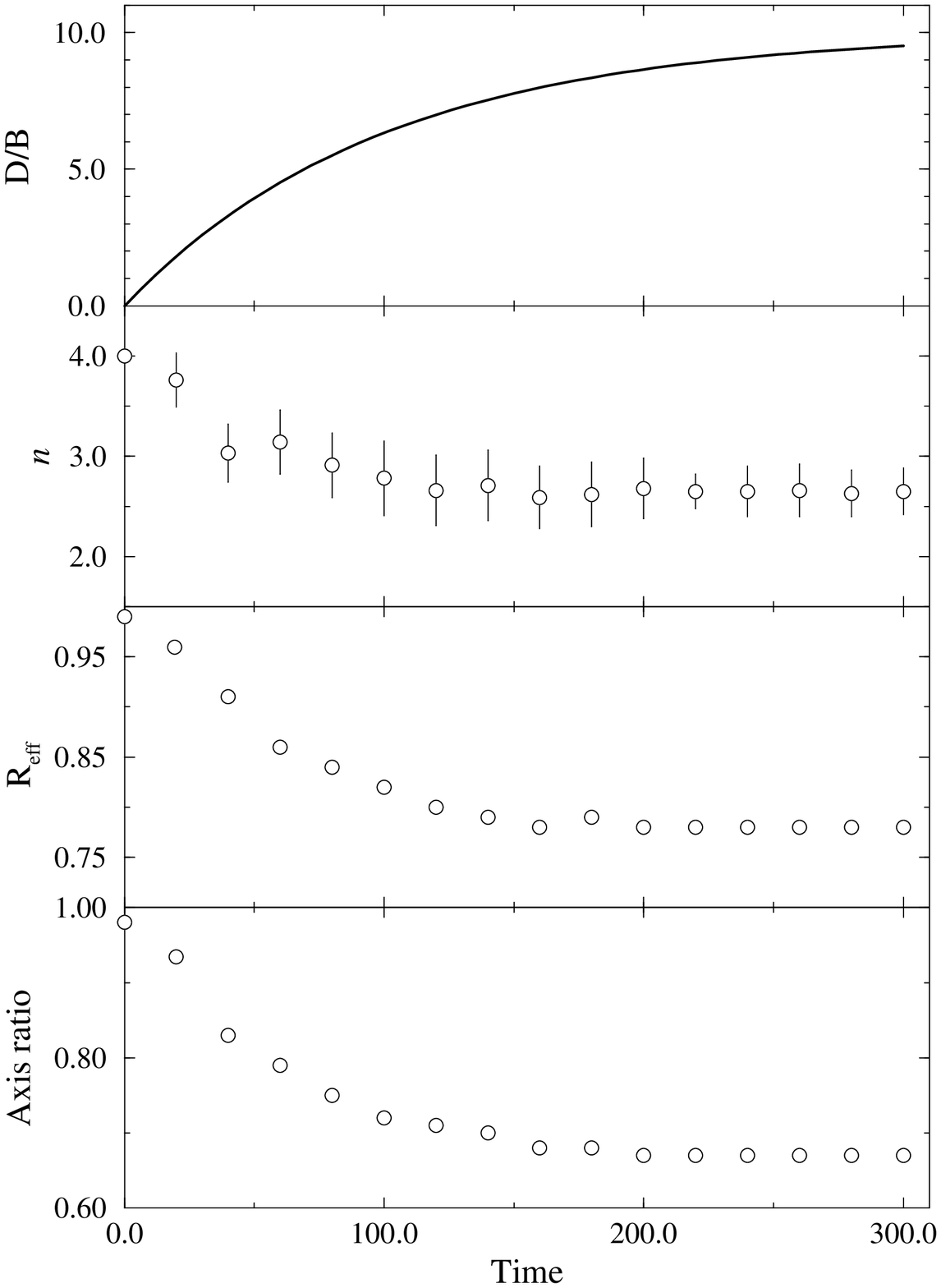}
\caption{{\bf Figure 2.} 
Time evolution of the basic bulge parameters during the simulation.
The increase of the disc mass is shown in the top panel. The index $n$ and the
axis ratio are determined using all the particles in the radial range
$0.1 < r < 5R_{\rm eff}$.}
\endfigure

\beginfigure{3}
\epsfxsize=7.8cm \epsfbox{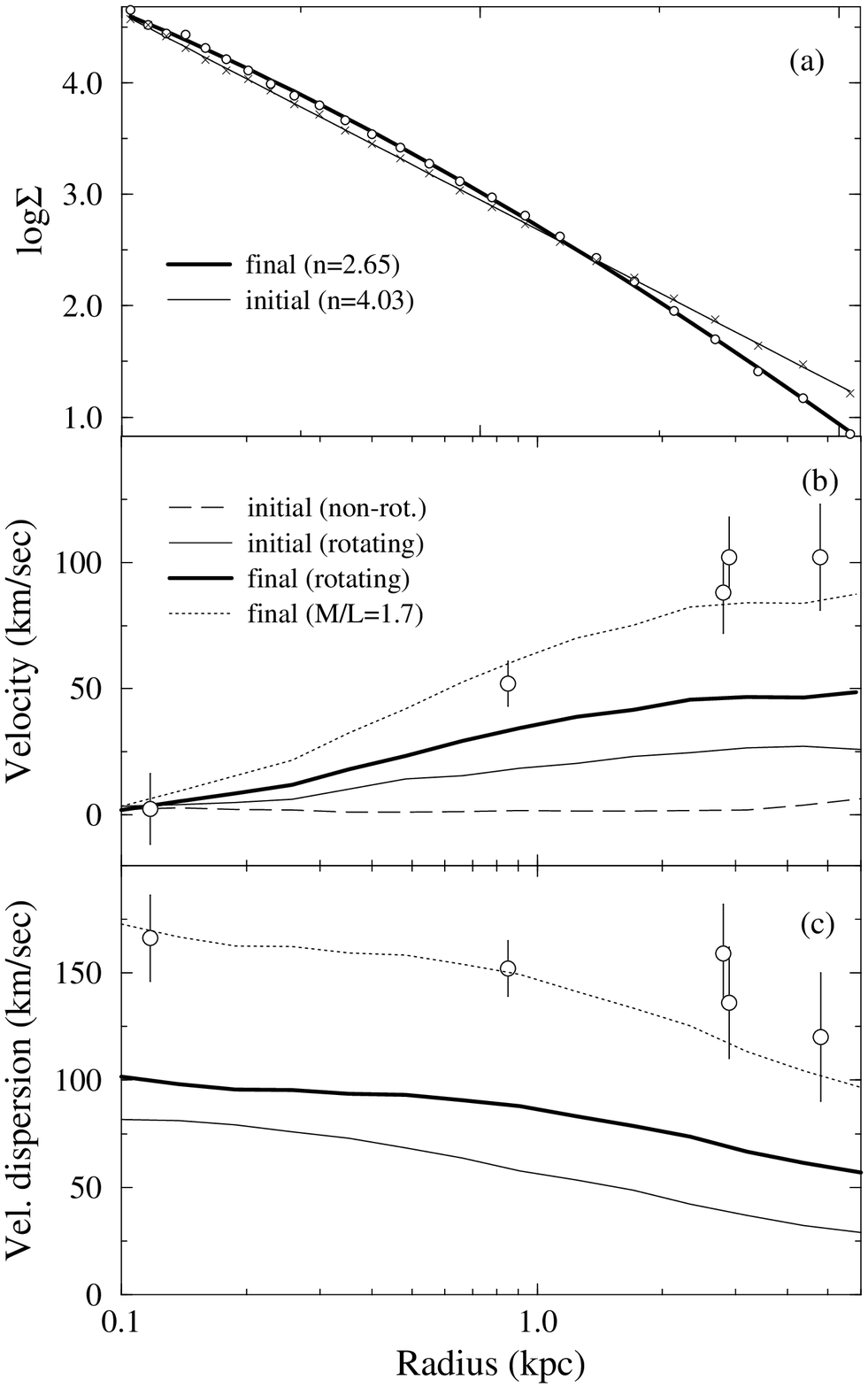}
\caption{{\bf Figure 3.} 
(a) The surface brightness profile of the bulge before (crosses) and after 
(circles) the growth of the disc. The errors are as a rule much smaller than 
the size of the data points ($\approx$ 0.03 in $\log\Sigma$). The best-fit laws are
given as thin and thick solid lines respectively. In this plot the radius axis
is scaled as \rqe, so that the de Vaucouleurs law is a straight line.
(b) The streaming velocities ($v_\phi$) before and after the disc.
(No model data points are given in this and the next plot, to avoid confusion.)
The dashed line represents the non-rotating models; the final curve is
identical to this one. The thin and thick solid lines represent the initial and 
final
state of the rotating models ($\gamma = 1$). The data points are the kinematics
of the bulge of NGC7814, and the dotted curve is the same as the final $v_\phi$
of the model (thick solid line), 
scaled to a total mass of 1.7 units. (c) The line-of-sight velocity dispersion of
the bulge. The line types are the same as in (b); the non-rotating models are
omitted here, as their velocity dispersion is almost identical to that of
the rotating ones.}
\endfigure
Both rotating and non-rotating models
are satisfactorily stable in all their parameters. 
The fluctuations in $n$, the axis ratio and 
the effective radius are all less than 2\% at all times, and for the entire
range of radii. A radial run of these parameters at the end of this isolated
evolution will be given later, along with the results of the disc models.
With the adopted values for the time step and the softening, the total energy 
as well as the angular and linear momentum are conserved to 0.1\% or better, 

In applying the disc field, the same technique is used as in Barnes \& White 
(1984). The potential of an infinitesimally thin disc with an exponential surface 
density, having a total mass \Md\ and a scalelength \hd\ is given by 
$$ \Psi_d(R,z) = -M_dG\,\int_0^\infty {J_0(kR)\over (1 + k^2h_d^2)^{3/2}}\,
                 e^{-k(z^2 + a^2)^{1/2}}dk \eqno\stepeq $$
where $a$ in the equation above is a softening parameter, used to 
avoid discontinuities in the vertical force. A value of $a = 0.05$ is used
throughout; the results do not show any dependency on $a$, for values between
0.1 and 0.01. The radial and vertical forces for \hd=1 and \Md=1 
are calculated on a 300$\times$300 logarithmic grid in $R$ and $z$. 
The field can then be easily scaled for other values of \hds and \Md.
The force at any position is obtained by linear interpolation in 
this grid. 

The increasing density of the disc is reproduced by making the
total mass increase as a function of time and approach asymptotically its
final value as
$$ M_d(t) = M_d(1 - \exp(t/\tau)) \eqno\stepeq $$
The timescale $\tau$ used for the growth of the disc is 100 units or 0.5 Gyr, in
general accordance with the disc formation timescales in the current picture
of galaxy formation (e.g. Fall and Efstathiou 1980).

Contour plots of the initial and the final equilibrium state of a ``generic"
bulge, with M = \Reffs = 1 and following initially a de Vaucouleurs profile, are
shown in Fig. 1, after the growth of a disc of mass $10^{11}$\solarm (B/D = 0.1) 
and a scalelength of 4 kpc. 
(Let it be stressed again here that the units used are conventional. The 
following results hold for any bulge-disc system with a B/D ratio of 0.1 and
a \hd/\Reffs ratio of 4.)
This is the final snapshot of the simulation, at 300 time units, although the
bulge has reached an almost steady state already at around t = 150. This can
be seen in Fig. 2, showing the evolution of the bulge parameters---$n$, \Reffs
and axis ratio $c/a$---as a function of time. Their evolution is analogous to
the applied disc force: they change rapidly in the beginning, and approach
asymptotically a final constant value after about 1.5 disc-growth timescales.
Fig. 3 shows the initial and final surface brightness profile, 
streaming velocities ($v_\phi$) and line-of-sight velocity dispersion as a function of
radius. The data points with the error bars in the $v_\phi$ and the
velocity dispersion plots are the kinematics of the bulge of NGC7814.  This
is one of the few galaxies with detailed bulge kinematics available (Kormendy
\& Illingworth 1982) as well as optical and near-infrared photometry (van der
Kruit \& Searle 1982 and Wainscoat et al. 1990 respectively). It is a spiral
galaxy of type Sab, with a bulge seen nearly edge-on and an effective radius
of $\approx$1.7 kpc, and thus almost ideal for modeling. Still, these data
are not to be compared directly to this particular simulation and are given
here only as an indication of whether or not the models are an adequate
representation of real bulges. The results in Figs. 1, 2 and 3 can be summarized
as follows:

\beginlist

\item (i) The bulge is flattened, as expected. Seen edge-on, the minor to 
major axis ratio (c/a) decreases going
outward from the center and its final value (measured using all the points 
within 5\Reff) is 0.65. The effective radius has decreased to 0.78, from
an initial value of 1.0. There is no significant difference in these parameters 
between the isotropic and the rotating model, although the outer isophotes of 
the rotating model (not shown) appear to be less discy than the ones of the 
isotropic; i.e. the rotating bulge is slightly less responsive in this respect.

\item (ii) The surface brightness profile has indeed steepened in the outer parts. The
initial value of the index $n$ was 4.03 (de Vaucouleurs law), and the final
value is $n_{iso} = 2.65 \pm 0.25$. The value for the rotating model 
is 
practically identical, $n_{rot} = 2.66 \pm 0.23$. 
These values are measured in the radial
range $0.1 < r < 5R_{\rm eff}$, somewhat larger than the range used to find the
\n values in the observations of APB95. The dependence of the value of \n 
on the range of points used, is discussed later. 

\item (iii) The maximum value of $v_\phi$ has increased by about 30\%, and the
velocity dispersion has increased also by a similar amount. For the rotating
model (and, in the case of velocity dispersion also for the isotropic one) both
these profiles are quite similar in shape with the ones of NGC7814. In fact,
all that is needed to match the observed velocities and velocity dispersions 
almost exactly, is a simple
scaling of the total mass of the model to $1.7\times10^{10}$\solarm, i.e. an
increase by a factor of 1.7. (dotted line in Fig. 3b and 3c). The $v_\phi$
curve falls slightly short of that of NGC7814 at the last point, but this can
be explained by the larger effective radius of the bulge of this galaxy (1.7 as
compared to 0.8 in the model).  This scaling is not intended to
reproduce exactly the kinematics of NGC7814; it only serves to show that
assuming a reasonable value for the mass, the kinematical properties of the
model are similar to those observed.

The effects on the kinematics will not be considered any 
further. They have been extensively studied by BW84, and the main purpose of
this paper is to examine the effects on the surface brightness. Suffice it to
say that the kinematics of the models are satisfactorily compatible with the
observed properties of real bulges. Finally, in view of the very similar 
response of the isotropic and the rotating models in terms of surface 
brightness, structure and velocity dispersion, the remainder of the discussion
will be confined to the non-rotating models. The rotating models are admittedly 
more realistic, but at the price of many arbitrary assumptions concerning 
the form of the DF and the value of $\gamma$ adopted in the angular momentum 
term.  
\endlist

\subsection{Radial dependence of the parameters and effects of 
                  different disc growth modes }

It is important to establish how parameters such as \n and axis ratio of the 
models depend on the radial range of points used to determine them. Fig. 4
shows the run of $n$ and the axis ratio $c/a$ as a function of the radius 
within which they are determined, for the final state of the bulge and for two 
different combinations: an extended disc (the one of the previous section,
with a scalelength of 4 kpc) and a more compact one, with a scalelength of
1.5 kpc, with the same mass. As one would expect, the effect of the compact
disc is more localized in $n$, with a broad minimum between r=1.5 and r=3,
corresponding to 3 and 6 \Reffs respectively. 
The extended disc causes the opposite effect. The central parts of the bulge
are now relatively denser and hence more resistant to the field of the diffuse
disc. This results in a local maximum in $n$, which then declines and finally
levels off at around $r=4$, or 5\Reffs for this bulge. 

Considering these results, it is best to determine $n$ within
5\Reffs for all the bulges, as a best compromise between
extended and compact discs. This limit ensures that enough particles are included 
for a good 
signal-to-noise ratio and it is also quite comparable to that of actual observations,
since the
brightness limit of 18.5 Kmag/arcsec$^2$ used by APB95 to calculate the \n of 
their bulge profiles corresponds usually to 3-4\Reff. (The fact that a
surface brightness limit corresponds to a roughly constant multiple of the
effective radius is due to the differences in total luminosity between the
bulges; those with the smallest effective radii are also the less luminous)

With respect to the axis ratio, the two discs yield quite similar 
results. For consistency, the axis ratio is measured within 5\Reffs as
well---this radius usually includes about 75\% of the particles.

The dependence of $n$ on the inclination of the system and on the angular
resolution used to measure the surface brightness profile are shown in the last 
two panels of Fig. 4. The final snapshot of the bulge (with an $n$ of 2.65) is 
considered in this experiment.   
As the inclination increases, $n$ shows a slight tendency to decrease; 
the $n$ at an inclination of 90$^\circ$ is about 5\% smaller that the one measured
face-on. This effect, however, does not appear to be significant as it lies well
within the error bars, and similar behaviour is also observed in the initial---spherical---%
model used for this simulation.
The same applies to the trend of $n$ with the angular resolution.
Made to simulate observations of bulges that lie progressively further away,
(and using a linear grid in radius instead of a logarithmic one)
this experiment shows that even when the resolution element (a ``pixel",e.g.)
is half the effective radius, the $n$ of the profile can still be retrieved,
albeit with a larger uncertainty (provided of course that the bulge {\it can}
be traced out to 4 or 5 effective radii).

\beginfigure{4}
\epsfxsize=7.8cm \epsfbox {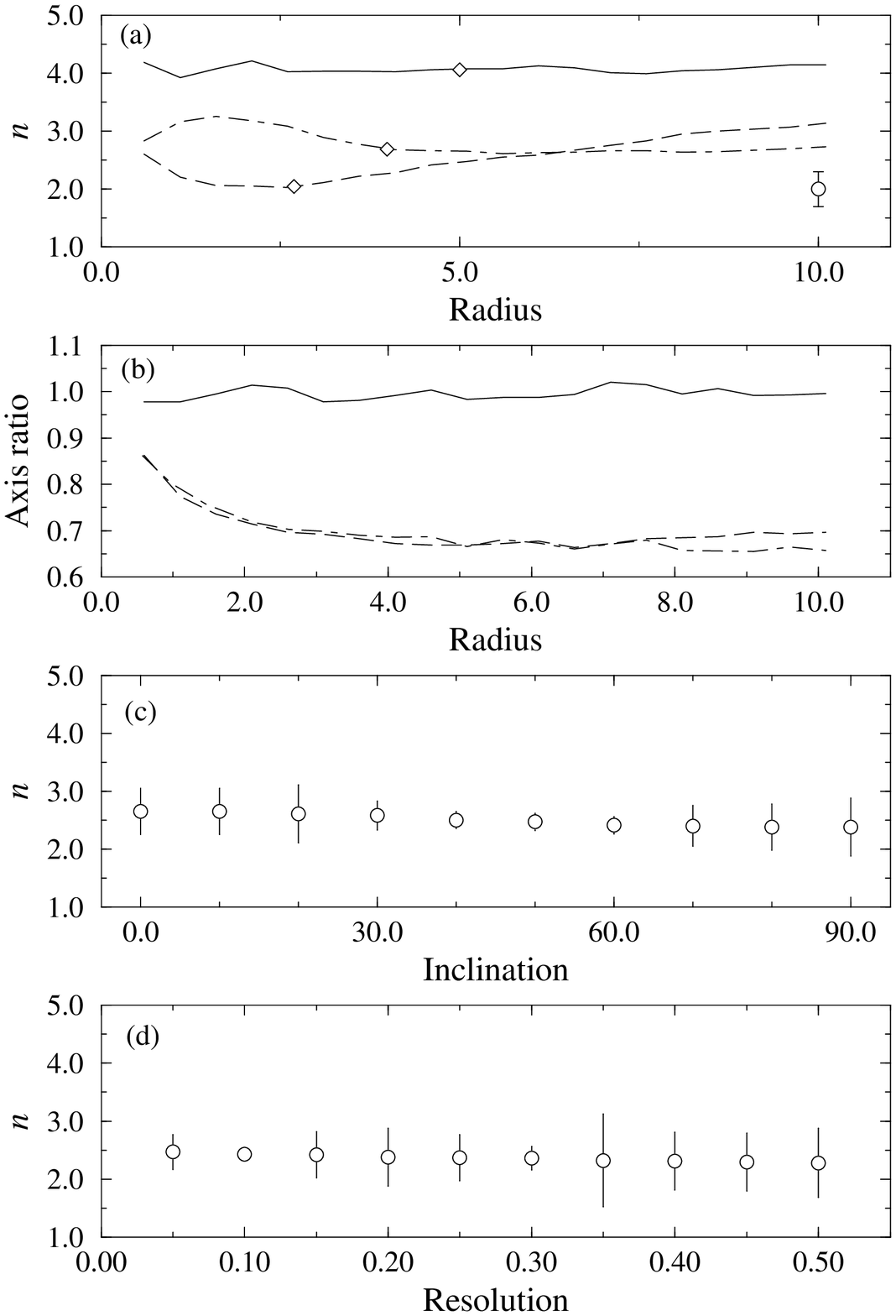}
\caption{{\bf Figure 4.} 
(a), (b) The values of $n$ and $\epsilon$ as a function of the radius
within which they are determined. Solid line: Final snapshot of model without a 
disc. Dashed line: disc with a scalelength of 1.5 kpc. Dot-dash line: Disc
with a scalelength of 4 kpc. Diamonds on top of each model mark the 5\Reff 
point. A typical error bar in $n$ is shown in the top panel.
(c) The dependence of the measured $n$ on the inclination of the galaxy, in degrees.
(d) The dependence of $n$ on the resolution of the grid (i.e. the ``pixel size'') used to
measure the surface brightness. The grid is linear in this case, and the 
resolution is in units of \Reff. Plots (c) and (d) refer to the simulation
with $h_d = 4$ kpc.}
\endfigure

\beginfigure{5}
\epsfxsize=7.8cm \epsfbox {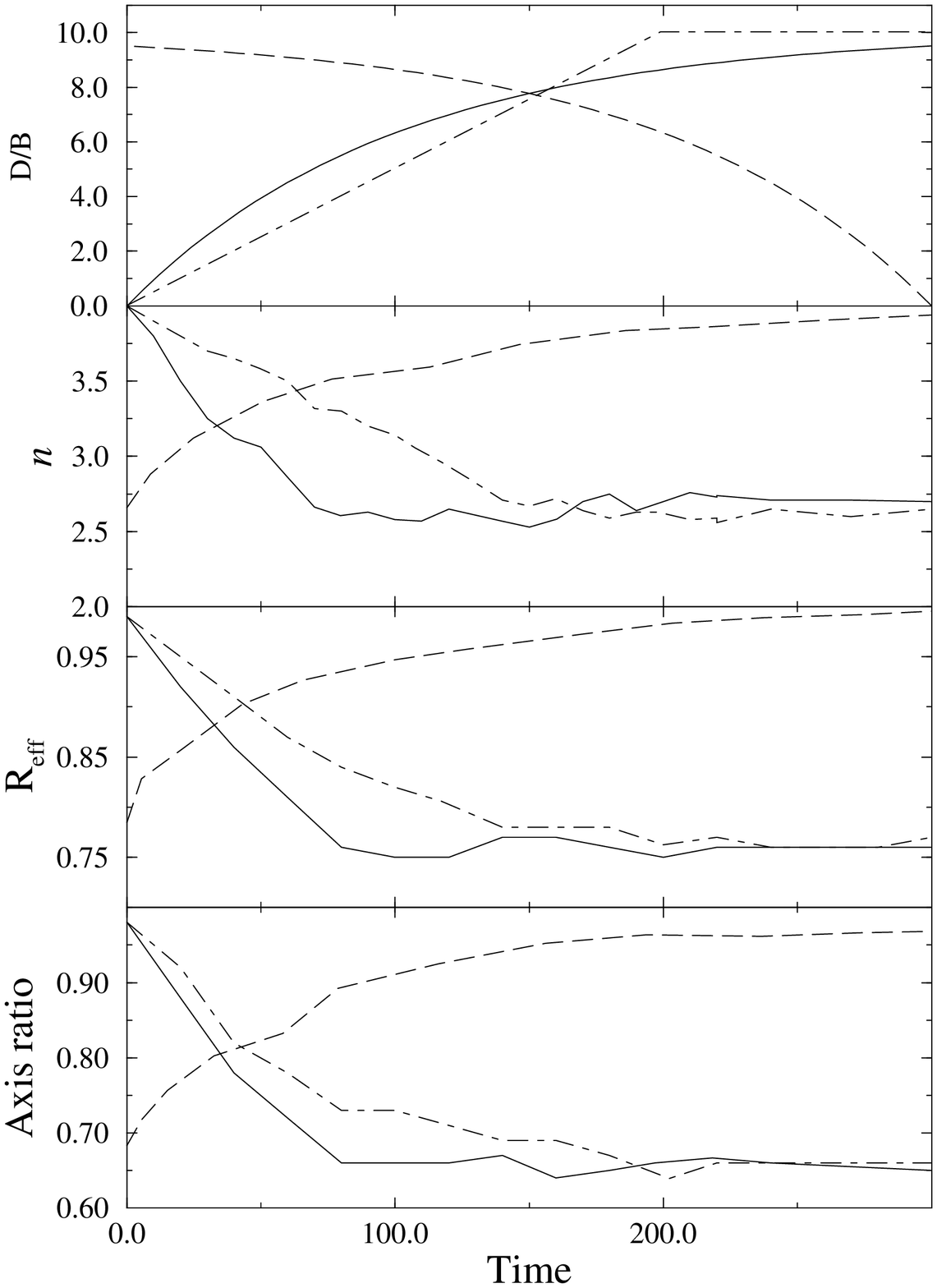}
\caption{{\bf Figure 5.} Time evolution of the bulge parameters---$n$,
effective radius and axis ratio---as the disc mass, shown in the top panel,
increases in different ways to the same final value:  Asymptotically with an
increasing disc scalelength (solid line in all panels) and linearly with time
(dot-dashed line in all panels). Notice the same final value of $n$, \Reffs and $c/a$ for these
two cases, despite the differences at $0 < t < 150$. The dashed line shows
the evolution of the bulge when the disc field is decreased back to zero.}
\endfigure

As mentioned also in BW84, the final state of the bulge should in principle
be independent of the way that the disc was formed, as long as the process is
adiabatic; in other words, if the timescale $\tau$ of the disc growth is large
compared to the dynamical time of the bulge. 
The dynamical time of a system with a mean density $\overline{\rho}$ is given by
$$t_{dyn} = \sqrt{{3\pi}\over{16\overline{\rho}}} 
                 = {\pi\over 2}\sqrt{{r^3}\over{M(r)}}. $$
For the present model, this gives a $t_{dyn}$ at the (initial) effective 
radius of approximately 5 time units, or $10^7$ years. 
Keeping always
within the limits of adiabaticity, this assumption is tested by growing the same 
disc as in the previous section in a timescale twice as long, $\tau = 200$. 
In addition, a linear instead of an exponential increase in mass is tried, as
well as a gradual increase of the scalelength \hds from 1.5 to 4 kpc, 
corresponding to a disc that grows from inside out, according e.g. to the
early models of Larson (1976).
These last two growth
modes and their results in terms of axis ratio, effective radius and $n$ are
shown in Fig. 5. The hypothesis of the independence of the
final state of the bulge on the details of the disc formation seems to be 
justified. The final equilibrium parameters of the bulge are the same within
their uncertainties, despite the significant differences at intermediate times. 
Finally, the standard test of removing
slowly the disc field from the final state of the bulge is done. As shown in Fig. 5, 
the bulge regains fully its original form and parameters, losing any ``memory" 
of the external potential.

\subsection{Dark Halo and non-adiabatic disc.}

For completeness, it is necessary to examine the effects of a dark
halo on the above results. It is customary to say that the halo dominates the
dynamics of the galaxy only in the outer parts; however, since the inferred
ratio of dark to visible matter varies a lot from galaxy to galaxy, it is
possible that a sufficiently massive halo can have drastic effects on the
response of the bulge to external fields. It has been more or less accepted 
in the last few years that dark halos in cosmological N-body simulations 
follow a $1/r$ density law in the
center (Dubinski \& Carlberg 1991, Salucci \& Persic 1996, Navarro et al. 1996).
 The density at 
large radii is still under debate, but this is not very important in this
study. Therefore, one may use the particularly simple form suggested by Hernquist (1990).
The density of this halo is
$$
\rho_H(r) = {M_H\over 2\pi}{\alpha\over r}{1\over (r + \alpha)^3} \eqno\stepeq $$
where $M_H$ and $\alpha$ are the mass and the scalelength of the halo.
The density falls off as $1/r$ at small radii and as $1/r^4$ in the outer parts.
This form of halo has been shown to give good fits to extended rotation
curves of spirals (Sanders \& Begeman 1994).
To make bulges that are in equilibrium inside such a halo, one simply adds the 
appropriate potential term in the DF, i.e. 
$$ \Psi(r) = \Psi_{bulge} + {GM_H\over r + \alpha} \eqno\stepeq $$ 
The model is constructed as before, and the halo potential is added to the
force calculation in the treecode. The rest of the procedure is the same, and 
the bulge-disc  configuration is the same as in \S 3.1. The system is 
evolved for three different 
values of the mass of the halo, namely for a total dark-to-visible mass ratio 
of 5, 10 and 30. The value of $\alpha$ is kept fixed to 70 kpc, according to
the $\alpha-M_{disc}$ relation found by Sanders \& Begeman (1994). The 
parameters of the bulge at the end of the simulation are given in Table 5. 
\begintable{2}
\caption{{\bf Table 2.} The effect of halos of different mass.}
{\halign{
\rm#\hfil&
\hskip 30pt\rm\hfil#&
\hskip 30pt\rm\hfil#&
\hskip 40pt\rm\hfil#\cr
$M_{halo}$ & \Reff & $c/a$ & \n ($\Delta n$) \cr
(1) & (2) & (3) & (4) \cr
\noalign{\vskip13pt}
  0.   &  0.75  &  0.65 &   2.65 (0.25) \cr
  50.  &  0.79  &  0.71 &   2.70 (0.17) \cr
  100. &  0.86  &  0.77 &   2.87 (0.22) \cr
  300. &  0.95  &  0.89 &   3.24 (0.31) \cr
}}
\tabletext {{\bf Columns in Table 2:} (1) Mass of the halo in $10^{10}$\solarm;
(2) Effective radius of the final bulge; (3) Axis ratio of the bulge at 5\Reff;
(4) \n of the bulge at 5\Reff. The initial mass and \Reff of the bulge are
$1 \times 10^{10}$\solarm and 1 kpc respectively, while the disc has 
$M_d = 10\times 10^{10}$ and ${\rm h}_d = 4$ kpc. The bulge consists of 32768
particles.}
\endtable

It is obvious that the halo, depending on its mass, can have significant
effects of the final outcome. The deepening of the potential well (reflected
on the much higher bulge velocity dispersion) offers additional support to the
bulge, making it less responsive to the field of the disc. At the limit
where the halo is extremely massive, the bulge can become completely 
insensitive to the disc field. The current estimates for the ratio of
dark to visible mass for normal spirals within the optical radius, lie around
a value of 1.0 or less (Salucci \& Persic 1996); for the halo used here this 
implies a total mass of 10 times the luminous mass, or 100 in the units of 
Table 2. For these values, the
results for the bulge are not significantly changed, and our previous analysis
is still valid. It should be stressed here, however, that this halo is not 
``live"; it is represented only by a potential. A real halo would also respond 
to the potential of the disc, and this could have an effect on the final 
result that cannot be readily anticipated. The response of fully realized
bulge-halo systems is deferred for future work.

A final test, is to grow the disc non-adiabatically. Using always the same 
configuration as in \S 3.1, the disc is grown in a timescale of just
1 time unit, much smaller than the dynamical time in almost every part of the
bulge. There is no reason to believe that such a rapid disc growth can
actually happen; this is done for the sake of completeness. The effective
radius and the axis ratio of the bulge change almost as much as in the adiabatic
limit, but this does not apply to the surface brightness profile. It remains relatively
close to an \rq law, showing that the adiabaticity is a necessary condition
for this mechanism to work. Some additional test runs showed that a rough
time limit for the non-adiabatic domain is 10 time units (0.05 Gyr). If the
disc grows in a timescale larger than this, the results in $n$ are the same as
those in \S 3.1.


\section{Comparison with observations}

\subsection{The parameter space of B/D and h$_{\rm d}$/R$_{\rm eff}$}

We now consider the effects of disc fields of different strength on 
the bulge parameters. 
In order to have a good estimate of the effects of the disc, one should 
explore as large a region of the parameter space as possible. Discs
of mass 0.5, 1, 4, 10 and 20 times the mass of the bulge are applied, the smallest ones
corresponding to S0 galaxies (B/D = 2) and the largest ones (B/D = 0.05) to
Sc spirals. For each one of these B/D ratios, scalelengths of 
1.5, 2, 3, 4, 6 and 10 kpc are tried, or 30 combinations in total. It is necessary to
take these different values for the scalelength, since, as shown in \S 3.2
a more compact disc has a bigger effect on the bulge than a larger one with
the same mass. A few runs are done also with a disc mass of 100, for 
completeness in the extreme low B/D regime. In all the runs the 
disc is grown exponentially as in eq. (7), with a timescale  $\tau$ of 100 
units. As in Fig. 2, the bulge parameters change quite rapidly initially, 
reach their final values at around $t = 150$ and remain constant thereafter. 

\beginfigure{6}
\epsfxsize=7.8cm \epsfbox {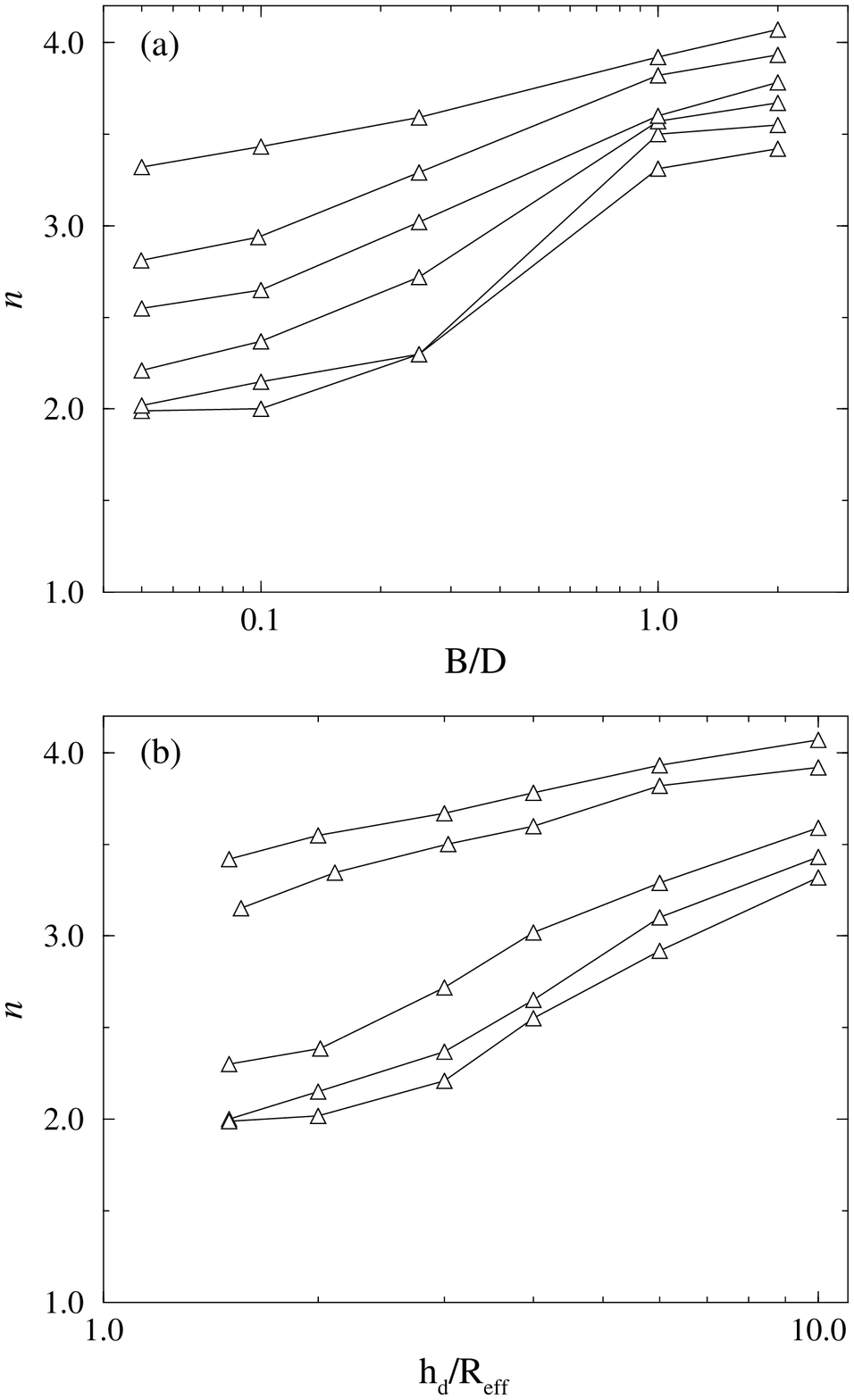}
\caption{{\bf Figure 6.} 
(a) The final values of $n$ as a function of the B/D ratio, for different
values of the scalelength ratio. From top to bottom, \hd/\Reffs = 10, 6, 4, 
3, 2 and 1.5.
(b) Final values of $n$ as a function of \hd/\Reff, for different B/D ratios.
From top to bottom, B/D = 2, 1, 0.25, 0.10, 0.05.}
\endfigure

\beginfigure{7}
\epsfxsize=7.8cm \epsfbox {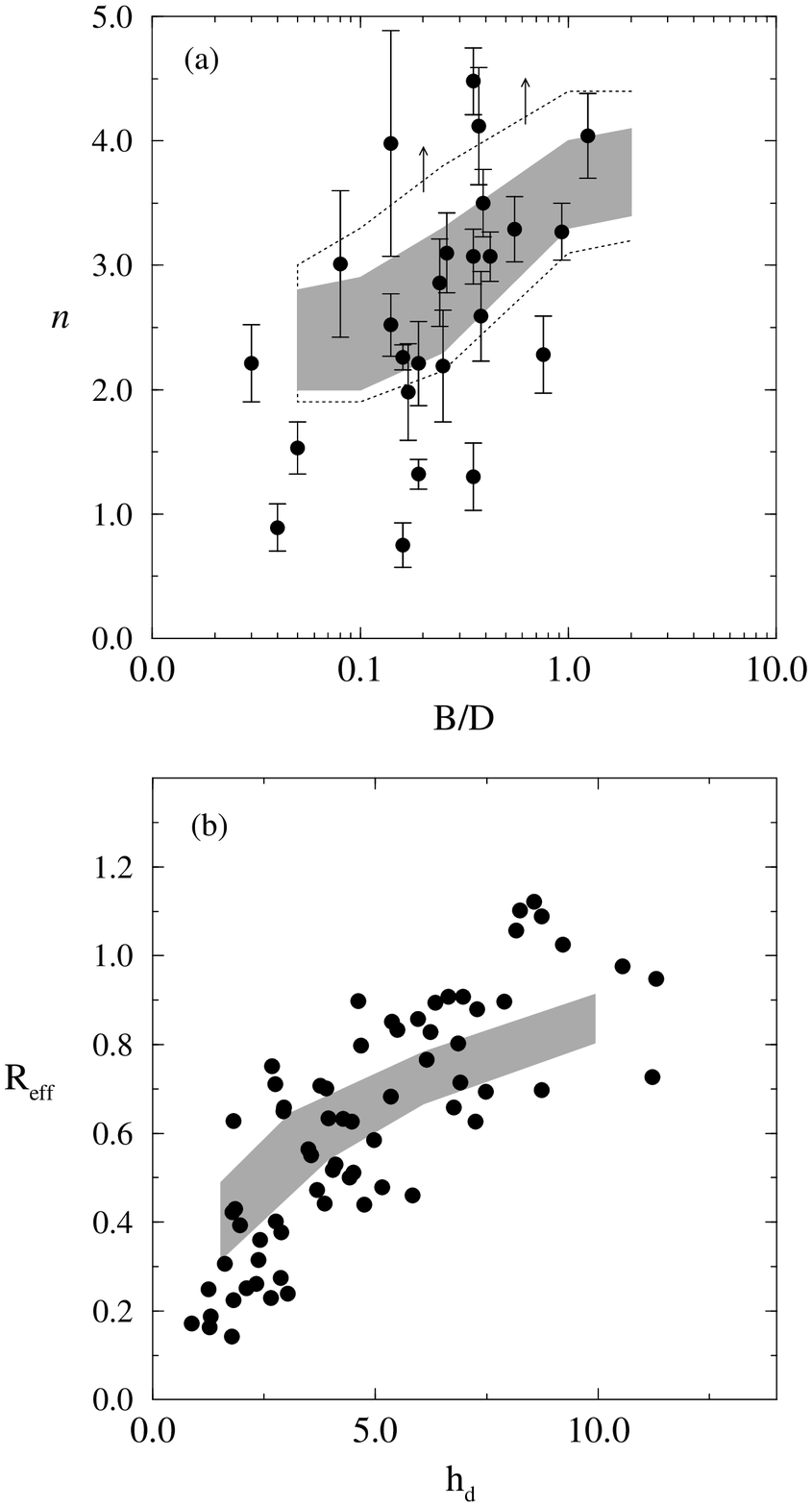}
\caption{{\bf Figure 7.} (a) The results of the simulations (shaded area) 
from Fig. 6a, shown on top of the observed trend. The dotted lines show the 
borders of the area if the errors in $n$ are taken into account.
The data points are the bulges of APB95. Notice the 
``rounding off" and saturation of the trend in the simulations at around $n=2$.
The arrows at the top show the way that the shaded area would expand if we
considered higher M/L ratios for the bulge.
(b) The relation between the disc scalelength and the bulge effective radius 
that results from the simulations (shaded area) shown on top of the results of 
de Jong (1995).} 
\endfigure

\beginfigure*{8}
 \epsfxsize=16cm \epsfbox {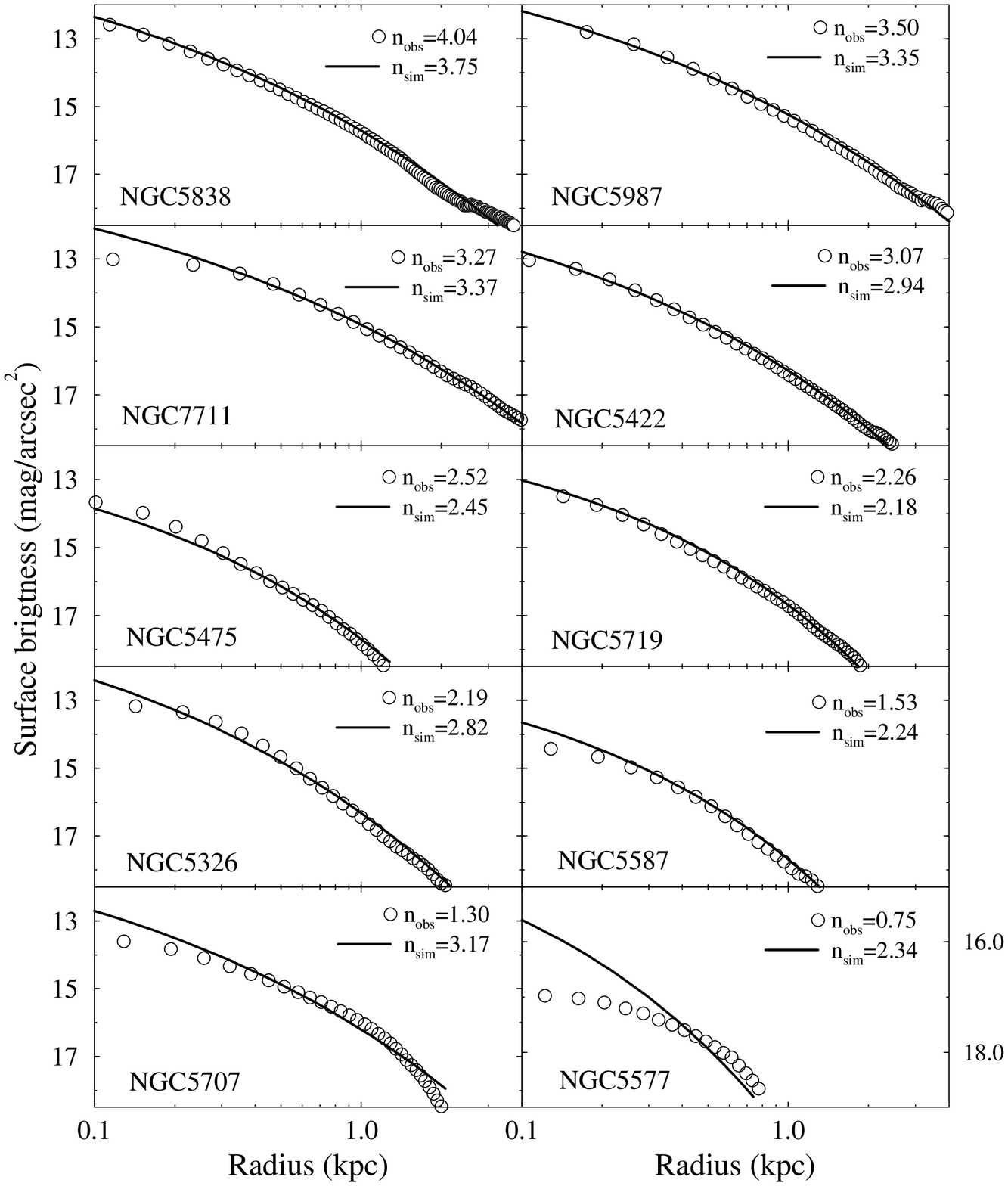}
 \caption{{\bf Figure 8.} 
 The observed bulge profiles (circles) and the results of the simulations 
 (solid line). The discrepancy in the central region of NGC7711 is due to
 seeing effects. The magnitude scale of NGC5577 is different (this bulge is much
 fainter) and is given on the left side of the corresponding panel.}
\endfigure

\begintable{3}
\caption{{\bf Table 3.} Results for a grid of parameters.}
{\halign{
\rm#\hfil&
\hskip 15pt\rm\hfil#&
\hskip 15pt\rm\hfil#&
\hskip 15pt\rm\hfil#&
\hskip 25pt\rm\hfil#\cr
B/D & \hd/\Reff & $c/a$ & \Reff & \n ($\Delta n$) \cr
\noalign{\vskip13pt}
  2.0  & 1.5  & 0.94  &  0.85   & 3.42 (0.24)   \cr
       & 2.0  & 0.94  &  0.90   & 3.55 (0.23)   \cr
       & 3.0  & 0.96  &  0.93   & 3.67 (0.25)   \cr
       & 4.0  & 0.97  &  0.94   & 3.78 (0.29)   \cr
       & 6.0  & 0.98  &  0.94   & 3.93 (0.30)   \cr
       & 10.0 & 0.98  &  0.94   & 4.07 (0.27)   \cr
  1.0  & 1.5  & 0.84  &  0.83   & 3.31 (0.28)   \cr
       & 2.0  & 0.84  &  0.87   & 3.50 (0.24)   \cr
       & 3.0  & 0.86  &  0.91   & 3.57 (0.21)   \cr
       & 4.0  & 0.90  &  0.92   & 3.60 (0.24)   \cr
       & 6.0  & 0.92  &  0.93   & 3.82 (0.42)   \cr
       & 10.0 & 0.92  &  0.94   & 3.92 (0.37)   \cr
  0.25 & 1.5  & 0.74  &  0.65   & 2.30 (0.14)   \cr
       & 2.0  & 0.75  &  0.72   & 2.30 (0.36)   \cr
       & 3.0  & 0.77  &  0.79   & 2.72 (0.50)   \cr
       & 4.0  & 0.79  &  0.85   & 3.02 (0.48)   \cr
       & 6.0  & 0.80  &  0.89   & 3.19 (0.55)   \cr
       & 10.0 & 0.82  &  0.91   & 3.59 (0.47)   \cr
  0.10 & 1.5  & 0.66  &  0.51   & 2.00 (0.12)   \cr
       & 2.0  & 0.66  &  0.63   & 2.15 (0.20)   \cr
       & 3.0  & 0.66  &  0.67   & 2.37 (0.38)   \cr
       & 4.0  & 0.68  &  0.76   & 2.65 (0.43)   \cr
       & 6.0  & 0.70  &  0.80   & 2.93 (0.40)   \cr
       & 10.0 & 0.72  &  0.91   & 3.43 (0.27)   \cr
  0.05 & 1.5  & 0.62  &  0.46   & 1.99 (0.10)   \cr
       & 2.0  & 0.62  &  0.52   & 2.02 (0.17)   \cr
       & 3.0  & 0.64  &  0.65   & 2.21 (0.19)   \cr
       & 4.0  & 0.64  &  0.72   & 2.55 (0.20)   \cr
       & 6.0  & 0.66  &  0.78   & 2.82 (0.23)   \cr
       & 10.0 & 0.66  &  0.87   & 3.32 (0.27)   \cr
}}
\tabletext {{\bf Notes on Table 3:} The axis ratio and \n are measured
within 5\Reff. The models consist of 16384 particles; this causes larger
errors, and some slight differences with e.g. the results in \S 3.1, with
double the number of particles. No dark halo is included in these simulations.}
\endtable

The final bulge parameters from all the runs are listed in Table 3. 
The values of \n at the end of the simulation range from around 3.8 down to 2.0 and the largest effects 
are caused by more massive and more concentrated discs; this constitutes the two 
main results of this study. The index \n does indeed decrease with
decreasing B/D ratio, as observed by APB95. However, the values of $n$
found here do not span the entire range observed. There are no bulges with $n < 2$. 
This can be seen in Fig. 6a, where $n$ is plotted as a function of the B/D
ratio, for the different values of \hd/\Reff. The $n$-B/D curves do not 
continue down to $n$=1 for more massive discs, but change slope and flatten out 
at $n=2$; this mechanism for changing
the surface brightness profile ``saturates", and exponential
bulges are not produced. This saturation is best illustrated if we
consider the results for discs having a mass of 10$\times$\tentensol and 
20$\times$\tentensol,
both with a scalelength of 1.5 kpc from Table 3. Despite the factor of 2
difference in mass, the final \n of the bulge is exactly the same. This is also
seen in Fig. 6b, where $n$ is plotted as a function of \hd/\Reffs for the
different values of B/D. 

In Fig. 7a, the \n-B/D relation from Fig. 6a is plotted on top of the observed relation
found by APB95 for the galaxies of their sample. There are 27 galaxies shown
in Fig. 7a. Three galaxies from the sample of APB95 with an $n$ larger than
4.5 have been omitted, since they have not been really included in the
modeling; an \rq law is assumed as initial state. The results of Fig. 6a are
given here as a shaded area. The lower boundary of this area is the $n$-B/D
curve for \hd/\Reffs=1.5, and the upper boundary is the $n$-B/D curve for
\hd/\Reffs=6; these are the values of \hd/\Reffs spanned by the galaxies of
APB95.

For $n > 2$, most of the galaxies lie inside this area and the
slopes of the two relations are similar, or, in any case, compatible. 
As expected after the results of Fig. 6, however, the 5 galaxies with $n < 2$, 
comprising 16\% of the diameter limited sample of APB95, lie clearly outside 
the simulation results; disc growth fails to account for the surface
brightness of these bulges. Some possible causes for this are discussed in 
Section 5. It should be noted here that a significant source of error   
is the implicit assumption in the 
simulations that the mass-to-light ratios of bulge and disc are the same.
This is often not the case, as has been shown in decompositions of rotation
curves. The M/L ratio of the bulge can be anything from half
of that of the disc for Sa spirals to 4 times bigger for later types 
(Broeils 1992, Kent 1986, 1987). A bigger M/L ratio would push the upper 
boundary of the area predicted by the simulations upward, as shown by the arrows. 
The lower boundary, however, is not affected by this and 
galaxies with $n < 2$ will always lie outside the predicted relation. 
A more direct demonstration of the above results is given
in Fig. 8, where a reproduction of the observed light profile of certain 
bulges from the sample of APB95 is attempted. 10 galaxies are selected, on the
basis of smoothness of the profile, reliability of the decomposition and error
in $n$. In addition, they should span as big a range in $n$ as possible.
A simulation is then run using the observed B/D ratio, and tuning the initial
\hd/\Reffs ratio so that the final effective radius is as close as possible to
the one observed. The results are given in Table 4, and plotted in Fig. 8.
To make a comparison possible,
the profiles from the simulations have been normalized to the same effective
surface brightness as the corresponding galaxy.
The conclusion is the same as above. The bulge profiles with
$n \ga 2$ can be reproduced very well; within the errors, the $n$ predicted by 
the simulations agree with the $n$ of the actual bulge profile for 6 galaxies
out of 10. For the remaining 4 galaxies with $n \la 2.2$ the discrepancy between
simulation and observed profile is greater as $n$ gets smaller.

\begintable{4}
\caption{{\bf Table 4.} Results of simulations of certain bulge profiles.}
{\halign{
\rm#\hfil&
\hskip 13pt\rm\hfil#&
\hskip 11pt\rm\hfil#&
\hskip 11pt\rm\hfil#&
\hskip 11pt\rm\hfil#&
\hskip 11pt\rm\hfil#&
\hskip 11pt\rm\hfil#&
\hskip 11pt\rm\hfil#\cr
NGC & B/D & \hd/\Reffs &$n_{obs}$&$\Delta n_{obs}$&$n_{sim}$&$\Delta n_{sim}$ \cr
\noalign{\vskip13pt}
5838  & 1.25 & 2.4 & 4.04 & 0.34 & 3.75 & 0.25 \cr
5987  & 0.39 & 2.7 & 3.50 & 0.27 & 3.35 & 0.25 \cr
7711 & 0.93 & 2.0 & 3.37 & 0.28 & 3.27 & 0.09 \cr
5422  & 0.42 & 2.9 & 3.07 & 0.20 & 2.94 & 0.50 \cr
5475  & 0.14 & 3.6 & 2.52 & 0.46 & 2.45 & 0.13 \cr
5719  & 0.16 & 2.7 & 2.26 & 0.10 & 2.18 & 0.11 \cr
5326  & 0.25 & 4.2 & 2.19 & 0.45 & 2.82 & 0.20 \cr
5587  & 0.05 & 3.9 & 1.53 & 0.21 & 2.24 & 0.13 \cr
5707  & 0.35 & 3.7 & 1.30 & 0.27 & 3.17 & 0.21 \cr
5577  & 0.16 & 3.3 & 0.75 & 0.59 & 2.34 & 0.11 \cr
}}
\endtable

Another aspect of these simulations that can be checked against observations
is the predicted relation between the final effective radius of the bulge and
the scalelength of the disc. According to the results in Table 3, if a 
standard initial bulge is assumed, then for a given B/D ratio the effective 
radius should be smaller for smaller \hd. The \hd-\Reffs relation in the K-band
found by
de Jong (1995) is plotted in Fig. 7b. Approximately 60 galaxies are included
here, with morphological types between 3 and 6 and B/D ratio between 0.01 and
0.15. The results of this paper are plotted again in the form of a shaded 
area. The lower boundary of this area is the predicted \hd-\Reffs relation for a
B/D ratio of 0.01 and the upper boundary is the same relation for a B/D ratio of 0.10.
While the slope of the two relations is not exactly the
same, the general trend is reproduced quite well, especially for the larger
bulges. A power law fit to the whole set of data gives
$$ \log{R_{\rm eff}^{\rm obs}} = 0.75\log{h_{\rm d}^{\rm obs}} - 1.62 $$
while for the galaxies with \Reffs $> 0.4$ the relation is 
$$ \log{R_{\rm eff}^{\rm >0.4}} = 0.40\log{h_{\rm d}} - 1.02 $$
The same fit on the simulation results gives
$$ \log{R_{\rm eff}^{\rm sim}} = 0.40\log{h_{\rm d}^{\rm sim}} - 1.07 $$
The error in the slope for the \Reffs $> 0.4$ bulges is $\pm$0.07, 
and the error in the simulation data is $\pm$0.05; the two relations 
agree very well. It should be noted here that, while the results in Fig. 7b 
assume that all the bulges have an
initial effective radius of 1 kpc, the relation would have the same form
for any ``flat" spectrum of initial effective radii.
The correlation between disc scalelength and bulge effective radius has been 
used, without firm theoretical justification, by
de Jong (1995) and Courteau et al. (1996) as a basic argument in favour of
the ``secular evolution" origin for bulges. It is shown here that a relation
of this form 
arises quite naturally if the bulge was formed before the disc. Again, as 
noted above, the smaller bulges, where the exponential profiles are
common, deviate from the predicted relation. 

\subsection{Different initial bulge profiles.}

We see that it is not possible in these simulations to obtain an exponential profile for the
bulge by growing a disc around an initially \rq system. There is no guarantee, 
however, that the initial bulge follows exactly an \rq law. For example, the 
initial profile might be already close to an \rh law, in which case it
would be seemingly easier for the disc to push it toward an exponential shape.
Here this hypothesis is tested by applying disc fields to initial bulges 
following \rh and exponential laws. The models are made as described in
\S 2.1, in the same way as the \rq law models. The results, for systems
having the same number of particles, total mass and effective radius as the
template \rq bulge of \S 3.1 are shown in Fig. 9 and listed in Table 5. 
\begintable{5}
\caption{{\bf Table 5.} Various initial bulge profiles.}
{\halign{
\rm#\hfil&
\hskip 35pt\rm\hfil#&
\hskip 15pt\rm\hfil#&
\hskip 15pt\rm\hfil#&
\hskip 15pt\rm\hfil#\cr
Model & $n_{init}$ & $n_{fin} (\Delta n)$ & $R_{\rm eff}$ & $c/a$ \cr
\noalign{\vskip13pt}
  \rq law      &   4.00   &  2.65 (0.25)  &  0.78  & 0.68 \cr
  \rh law      &   2.00   &  1.48 (0.13)  &  0.78  & 0.70 \cr
  Exponential  &   1.00   &  0.89 (0.08)  &  0.81  & 0.70 \cr
}}
\tabletext {{\bf Notes on Table 5:} The initial \Reffs for all the bulges is 1 kpc. The
models consist of 32768 particles.}
\endtable
The results do not conform to the expectations.
For the same disc, the changes in effective radius and axis ratio are almost
identical, but the change in \n is smaller as the initial bulge is closer 
to exponential. The \n of a de Vaucouleurs bulge is reduced by almost 40\%,
that of an \rh bulge by 20\% and that of an exponential bulge just by 10\%.
The $n=1$ (exponential) bulge is the most sturdy against the growth of a disc.
This increasing difficulty of changing \n depending on its 
initial value can readily explain the saturation at $n=2$ that is seen
in the results of Section 4.1 and in Figure 6: Once close to $n=2$, the profile
becomes more resistant to an already weakened disc field, and $n=1$ profiles can
never be created. 

\beginfigure{9}
\epsfxsize=7.8cm \epsfbox {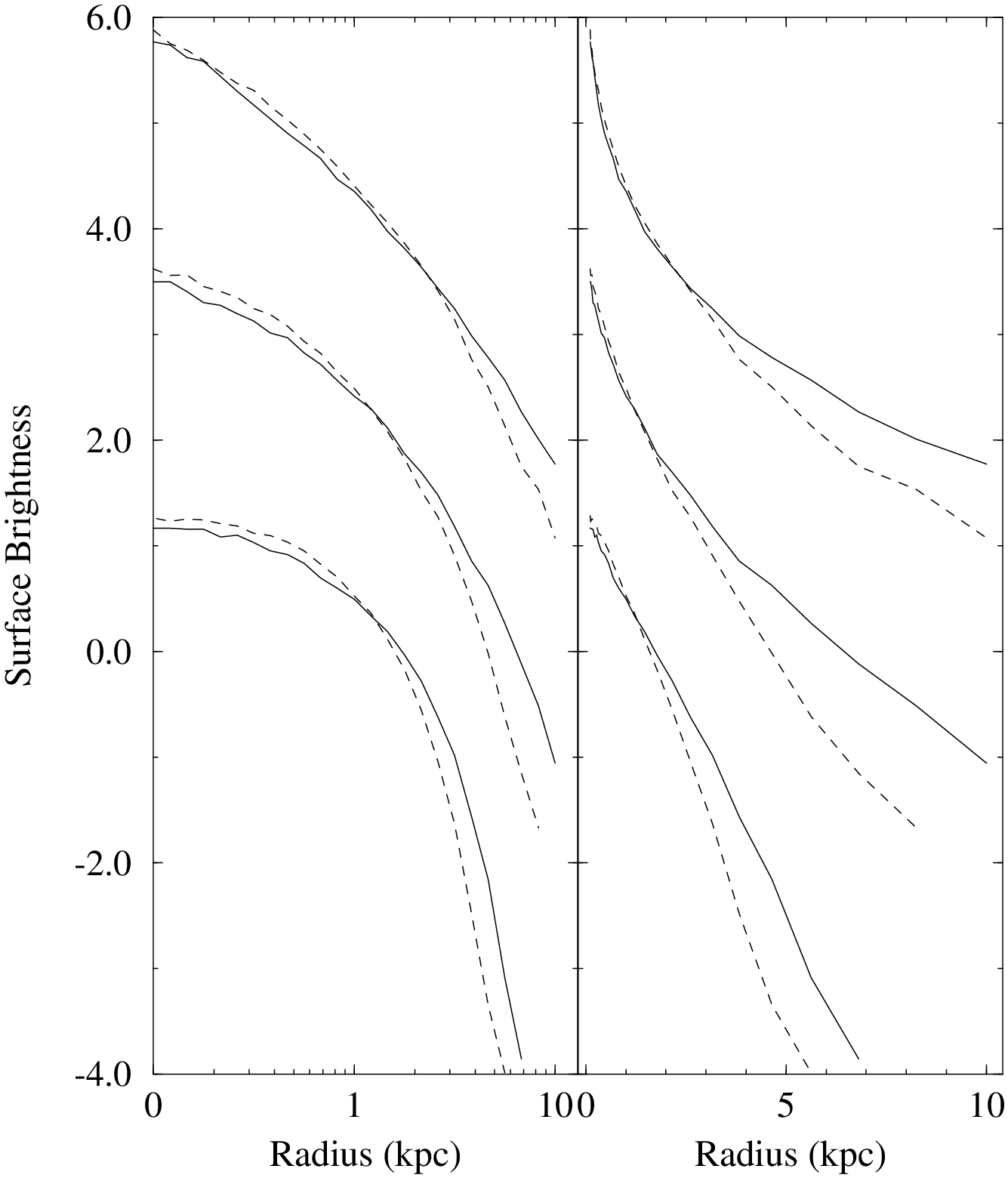}
\caption{{\bf Figure 9.} The initial (solid line) and final (dashed line) 
states of various density profiles, under the influence of the same disc. 
From top to bottom, \rq profile, \rh and exponential. (The profiles have been
shifted vertically for clarity.) We give the results
both in logarithmic (left panels) and in linear (right panels) scale, to 
show the differences at small and large radii respectively.} 
\endfigure

\section {Discussion}

Despite many simplifying assumptions, such as common M/L ratio for the bulge
and the disc, same luminous-to-dark-matter ratio for all the galaxies, and
exclusion of dissipative processes, the $n$-B/D relation is fairly
successfully reproduced for galaxies with $n > 2$. 
The \hd-\Reffs relation is also successfully reproduced, with the possible
exception of the smallest bulges. These two facts show that the
initial hypothesis may well be correct: These bulges existed before the disc
as \rq spheroids, and were later modified by the changing force field of the
disc to their present state. 

Bulges in the region $2 > n > 1$, however, remain unexplained.  If they were
created by the same processes as the early type bulges, i.e.  as \rq spheroids,
this drastic change in their profiles is unaccounted for.  The answer might
well be that, simply, they were not created by the same processes. As mentioned
in the introduction it has been shown that there exist other mechanisms, such
as the destruction of bars, that can create bulge-like entities in the centers
of spirals. Bars themselves can successfully play the role of bulges; in fact,
as has been shown in numerical simulations, a bar seen end-on displays an
exponential surface brightness profile (Pfenniger \& Friedli 1991) and this
fits very well into the whole scenario. The necessary dichotomy, though, in
origin and properties, remains somewhat disturbing. The properties of $n=1$
bulges, although deviant, are a continuation of those with $n > 2$; there is no
apparent bimodality neither in $n$, nor in any of the other parameters (see,
e.g., the fundamental plane of bulges and ellipticals in APB95).  A possible
way out of this dilemma is that the bars in late type spirals (appearing as
exponential bulges) could cover an already existing small, dissipationally
formed bulge; if the bar is formed on top of the bulge, the two can co-exist
(Combes \& Sanders 1981, Combes et al. 1990).  Actually, it can be argued that
when the bulge is small enough---and hence having a profile close to the \rh
law---it is {\it then} that the disc becomes unstable against bar formation.
Ostriker and Peebles (1973) pointed out that for the disc to be stable against
bar formation, the ratio $t$ of rotational energy to the total potential energy
has to be less than about 0.14.  Sellwood (1980) showed that the
Ostriker-Peebles criterion could be satisfied by the existence of a bulge
having at least 50\% of the mass of the disc; this percentage could be even as
low as 30\% (Berman and Mark 1979).  If this low percentage is correct, then
galaxies with a B/D ratio of less than 0.3 are candidates for bar formation,
and, as a result, for having ``bulges" with exponential profiles. In this way
the spectrum of $n$ for the central spheroids---be they old, dissipationally
formed bulges or not---would have the desired continuity. However, there are
many other possible scenarios:  \beginlist

\item First is that the initial bulges already span the whole
spectrum of $n$ values, from 4 (or even larger) down to 1. All of them would
then be affected by the disc: the highest values of $n$, depending on the
B/D ratio, would decrease, while the lowest values, as shown, will remain
almost unaffected. However, a mechanism that can produce this wide range of bulge
profiles at birth is still lacking.

\item Another possible choice is that all bulges are created as exponential
spheroids in the beginning, and then disrupted by minor mergers. In general,
merging processes produce \rq law systems (Barnes \& Hernquist 1992 and
references therein). As a
result, the bulges that have undergone more merging would be more massive,
more luminous, and have \rq law profiles. The undisturbed bulges would remain
exponential in profile and small in mass. Again, a mechanism that can produce
primordial exponential bulges is still to be found.

\item Finally, one should consider the possibility that $n=1$ bulges
do not exist at all; that they are in fact larger $n$ systems mistaken for exponential,
because of poor resolution, low signal-to-noise or disc contamination. 
This does not seem very probable. Exponential profiles have also been found 
in large and luminous bulges (e.g. NGC5707 and NGC7311, APB95), and they have
arisen in one- and two-dimensional bulge-disc decompositions (de Jong,
1995, Courteau et al. 1996), as well as in the decomposition of the inner parts
of rotation curves (Heraudeau et al., 1996). Moreover, it is shown 
in this paper (\S 3.2) that poor resolution cannot produce this effect.

It should be added here that the central regions of late type bulges remain largely
unexplored. We still do not know enough about the density and surface brightness profile
at radii smaller than about 0.05 kpc, and this could provide some better 
constraints on the possible scenarios mentioned above. The first HST data on a sample of 
S0 bulges (Phillips et al. 1996) indicate that there exists a large 
variety of profiles at the center; more data are clearly needed. 
\endlist

\section {Conclusions}

We try to reproduce the observed trend of the surface brightness profile of
bulges to become steeper in the outer parts for the later type spirals. It is
attempted to see whether or not this trend can be explained by the growth of a disc
of different mass around each bulge. If this is the case, the implication is that the
bulge was formed first, and we are looking at the imprint of the disc
formation on it. Equilibrium spherical models following an \rq law are built
to represent the initial bulge, and then the force field of a thin
exponential disc whose density increases with time is applied. 
It is found that

1. The index $n$ of Sersic's law $\Sigma_n(r) \propto \exp[-(r/r_0)^{1/n}] $
that describes the surface brightness profile is smaller
than the initial value of 4 (de Vaucouleurs' law) after the disc formation is 
complete. More massive
discs (relative to the bulge), or discs with a smaller scalelength lead to a 
smaller value of $n$, in agreement with the observations. The biggest part of 
the observed correlation between the scalelength of the disc and the effective
radius of the bulge is explained, and the slope of the
$n$ -- B/D ratio relationship is compatible with the one observed.

2. The decrease in the value of \n saturates at $n=2$. Bulge profiles that
are steeper than the \rh law cannot be produced by this mechanism. 
Spheroids are increasingly resistant to change in the slope of
their surface brightness profiles, as they become steeper in the outer parts 
(i.e. having smaller \n initially). The exponential profile ($n=1$) is a real 
barrier in this sense: It remains exponential, despite the changes in 
effective radius, axis ratio, velocity dispersion etc. 

These results support the idea that bulges of early type spirals were formed 
before the discs.
The systematic differences in the radial distribution of surface brightness for
these bulges are explained as the imprint of the formation of the disc,
later in the life of the galaxy. The genuine exponential bulges, though, if
they are real, are not explained by such a mechanism. These results suggest that they
were either formed as such---in which state they will remain, as shown here,
{\it ad infinitum}---or they are the result of secular evolution phenomena later in
the history of the galaxy, such
as bar formation and/or destruction.

\section*{Acknowledgements}

The author would like to thank Marc Balcells for his help in the beginning of
this project, Roelof de Jong for sending his data, and Reynier Peletier
and Andy Burkert for useful suggestions and discussions. Many thanks are due
to Bob Sanders for numerous readings of the manuscript, comments,
suggestions and encouragement. The softening of the distribution function was
suggested by Linda Sparke. The treecode was obtained by anonymous ftp from the
University of Hawaii, kindly made available by Joshua Barnes.
The simulations were done on the CRAY J90 of the Computer Center of the
University of Groningen.

%



\section*{References}
\beginrefs

\bibitem Andredakis Y.C., Peletier R.F., Balcells M., 1995, MNRAS, 275, 874 (APB95)
\bibitem Andredakis Y.C., Sanders R.H. 1994, MNRAS, 267, 283
\bibitem Antonov V. A., 1962, Vestnik Leningrad Univ., 7, 135 
\bibitem Balcells M., Peletier R.F., 1994, AJ, 107, 135
\bibitem Barnes J., Hernquist L., 1992, ARAA, 30, 705 
\bibitem Barnes J., Hut P., 1986, Nature, 324, 446
\bibitem Barnes J., White, S.D.M., 1984, MNRAS, 211, 753 (BW84)
\bibitem Baugh C.M., Cole S., Frenk C.S., 1996, MNRAS, 283, 1361
\bibitem Berman R.H., Mark J.W., 1979, A\&A, 77, 31
\bibitem Binney J., 1982, MNRAS, 200, 951
\bibitem Carney B., Latham D., Laird J., 1990, in Jarvis B., Terndrup D., eds, 
         ESO Proc., 'Bulges of Galaxies'
\bibitem Combes F., Debbasch F., Friedli D., Pfenniger D., 1990, A\&A, 233, 82 
\bibitem Combes F., Sanders R.H., 1981, A\&A, 96, 164
\bibitem Courteau S., de Jong R.S., Broeils A.H., 1996, ApJ, 457, L73
\bibitem Dehnen W., 1993, MNRAS, 265, 250
\bibitem de Jong R.S., 1995, Ph.D. Thesis, Univ. of Groningen
\bibitem Doremus J. P., Feix M. R., Baumann G., 1971, Phys. Rev. Lett., 26, 725
\bibitem Dubinski J., Carlberg R.G., 1991, ApJ, 378, 496
\bibitem Eggen O., Lynden-Bell D., Sandage A., 1962, ApJ 136, 748
\bibitem Fall M.J., Efstathiou G., 1980, MNRAS, 193, 189
\bibitem Frogel J.A., 1988, ARA\&A, 26, 51
\bibitem Heraudeau P., Simien F., Mamon G.A., 1996, in Minniti P., Rix H.-W.,
         eds, ESO Proc., `Spiral Galaxies in the Near-IR', Springer-Verlag 
         Berlin, p. 235
\bibitem Hernquist L., 1990 ApJ, 356, 359
\bibitem Jablonka P., Martin P., Arimoto N., 1996, AJ, 112, 1415
\bibitem Jarvis, B.J., Freeman, K.C., 1985, ApJ, 295,314 
\bibitem Katz N., 1991, ApJ, 368, 325 
\bibitem Kauffmann G., White S.D.M., Guiderdoni B., 1993, MNRAS, 264, 201 
\bibitem Kent S.M., 1986, AJ, 91, 1301
\bibitem Kent S.M., 1987, AJ, 93, 816
\bibitem Kent S.M., Dame T., Fazio G., 1991, ApJ, 378, 131
\bibitem Kormendy J., Illingworth G., 1982, ApJ, 256,460
\bibitem Kormendy J., 1992, in Dejonghe H., Habing H., eds, Proc. IAU Symp. 153, 
         Galactic Bulges. Kluwer, Dordrecht p. 209
\bibitem Kuijken K., Dubinski J., 1994, MNRAS, 269, 13
\bibitem Larson R. B., 1976, MNRAS, 176, 31
\bibitem Norman C.A., Sellwood J.A., Hasan H., 1996, ApJ, 462, 114
\bibitem Navarro J.F., Eke V.R.,  Frenk C.S., 1996, MNRAS, 283, L72 
\bibitem Navarro J.F., Frenk C.S., White S.D.M, 1996, ApJ, 462, 563
\bibitem Ortolani S., Renzini A., Gilmozzi R., Marconi G., Barbuy B., Bica E., \&
         Rich, R.M., 1995, Nature, 377, 701
\bibitem Ostriker J.P., Peebles P.J.E., 1973, ApJ, 186, 467
\bibitem Pfenniger D., Friedli D., 1991, A\&A, 252, 75
\bibitem Phillips A.C., Illingworth G.D., MacKenty J.W., Franx M., 1996, AJ, 
         111, 1566
\bibitem Salucci P., Persic M., 1997, in Persic M., Salucci P., eds, ASP Proc., 
         `Dark and Visible Matter in Galaxies', in press
\bibitem Sanders R.H., Begeman K.G., 1994, MNRAS, 266, 360
\bibitem Sellwood J.A., 1980, A\&A, 89, 296
\bibitem Sersic J.L., 1968, Atlas de galaxias australes. Observatorio Astronomico, Cordoba
\bibitem Shaw M., 1993, MNRAS, 261, 718
\bibitem Theis C., Hensler G., 1993, A\&A, 280, 85
\bibitem Tremaine S., et al., 1994, AJ, 107, 634
\bibitem van Albada T., 1982, MNRAS, 201, 939
\bibitem van der Kruit P.C., Searle L., 1982, A\&A, 110, 79
\bibitem Wainscoat R.I,, Hyland A.R., Freeman K.C., 1990, ApJ, 348, 85

\endrefs
\bye